\documentclass[12pt]{article}

\begin{document}
\begin{center}

{\bf{\Large  Regular Spherically Symmetric Interior Solution To Schwarzschid's Solution Which Satisfies The Weak Energy Conditions
}}

\vspace{1cm}

E. Kyriakopoulos\\
Department of Physics

National Technical University

157 80 Zografou, Athens, GREECE\\

E-mail: kyriakop@central.ntua.gr\\

Essay written for the Gravity Research Foundation 2016 Awards for Essays on Gravitation\\

Submitted February 26, 2016
\end{center}

\begin {abstract}

   We present a simple spherically symmetric and regular solution of Einstein's equations with two parameters $k$ and $M$, which matches                                                                                                                                  to Schwarzschild's solution, satisfies the weak energy conditions in the interior region and for small $r$ behaves like the de Sitter solution. Its energy density $\rho$ and its radial pressure $p_r$ satisfy the relation $\rho+p_r=0$. For some values of $k/M$ the solution does not have an event horizon and the event horizon of Schwarzschild's solution is inside the matching surface. Therefore it describes the formation of a gravitational soliton, which is shown to be stable. Gravitational solitons are related to dark matter. For the other values of $k/M$ it is a regular black hole solution.

PACS number(s): 04.20.-q,  04.20.Jb,  04.70Bw

\end {abstract}

In recent years several people tried to find regular black hole models \cite{Le1}-\cite{An1} as possible end products of gravitational collapse. The models can be divided into three classes:

First class: It contains models with no junction. The solution is continuous throughout the space-time. A general characteristic of these models is that the solution behaves for big $r$ like Schwarzschild's solution and for small $r$  like de Sitter's solution.. Several models of this class have been found \cite{Dy1}-\cite{Ha1}.

Second Class: The solutions of this class have two regions and a surface layer, i.e. thin shell, joining these regions. Usually a de Sitter core is joined with a Schwarzschild or Reissner-Nordstrom space. Regular black holes with thin shells of spacelike and timelike character have been found. Several papers of this class have been published \cite{Go1}-\cite{Ba2}.

Third Class: In this case we have two regions and a boundary surface joining them. Thin shell is not needed. Relatively little work has been done to find solutions of this class \cite{Ma1}-\cite{Co1}.

The metric we present in this paper, which is the metric of a model of the third class, depends on two parameters $k$ and $M$. It matches to the solution of Schwarzschild with gravitational radius $r_g=2 M$ on the surface $r=k$, giving a regular and simple interior solution to Schwarzschild's solution. For small $r$ the solution behaves like de Sitter's solution. It is anisotropic and satisfies the weak energy conditions in the interior region. It satisfies the relation $\rho+p_r=0$. The energy density $\rho $ is monotonically decreasing with $r$  and vanishes at the boundary surface.

The relation $g^{rr}=0$ is satisfied for three values of $x=k/M$.  Since for all values of $w=k/{6M}$ one of the values of $x$ is negative the solution has at most two event horizons. However if $w>256/729$ the two other values of $x$ are complex, which means that the solution does not have an event horizon. In this case the gravitational radius of Schwarzschild's solution and the matching surface $r=k$ satisfy the relation $r_g<k$, which means that the event horizon of Schwarzschild's solution is inside the matching surface. Therefore the combination of Schwarzschild's solution with the solution we have found does not have an event horizon. If $w<256/729$ the solution has two event horizons, which coincide  for $w=256/729$. We conclude that the regular solution we have found combined with Schwarzschild's solution describes the formation of a gravitational soliton in the first case and the formation of a regular black hole in the second case.

Consider the metric

\begin{equation}
ds^2=-(1-\frac{8 M r^2}{k^3}+\frac{6 M r^3}{k^4})dt^2+(1-\frac{8 M r^2}{k^3}+\frac{6 M r^3}{k^4})^{-1}dr^2+r^2(d\theta^2+sin^2\theta d\phi^2)
\label{1}
\end{equation}
and the metric of Schwarzschild's solution

\begin{equation}
ds^2=-(1-\frac{2 M}{r})dt^2+(1-\frac{2 M}{r})^{-1}dr^2+r^2(d\theta^2+sin^2\theta d\phi^2)
\label{2}
\end{equation}
We can easily show that the metric coefficients of expressions (\ref{1}) and (\ref{2}) are identical at $r=k$ and the same thing happens for the derivatives with respect to $r$ of these coefficients. Therefore the surface $r=k$ is a matching surface of the metrics of Eqs  (\ref{1}) and (\ref{2}).

The non zero components of the Ricci tensor coming from the metric of Eq. (\ref{1}) are the following \cite{Bo1}
\begin{equation}
R_{tt}=-\frac{12M(2 k-3 r)(k^4-8 k M r^2+6 M r^3)}{k^8},\>\>R_{rr}=\frac{12 M(2 k-3 r)}{k^4-8 k M r^2+6 M r^3)}
\label{3}
\end{equation}
\begin{equation}
R_{\theta\theta}=\frac{24 M(k-r)r^2}{k^4},\>\>\>\>R_{\phi\phi}=\frac{24M(k-r)r^2sin^2\theta}{k^4}
\label{4}
\end{equation}
From the above expressions we can calculate the eigenvalues $\lambda_\mu$, $\mu=t, r, \theta, \phi$ of the matrix $R^\mu_\nu$. We get
\begin{equation}
\lambda_t=\lambda_r=\frac{12 M(2 k-3 r)}{k^4},\>\>\>\>\>\>\>\>\lambda_\theta=\lambda_\phi=\frac{24 M(k-r)}{k^4}
\label{7}
\end{equation}
Also the Ricci scalar coming from Eqs  (\ref{3}) and (\ref{4}) is the following \cite{Bo1}
\begin{equation}
R=\frac{24 M(4 k-5 r)}{k^4}
\label{9}
\end{equation}

The eigenvalues $v_\mu=\lambda_\mu-R/2$ of Einstein's tensor $R^\mu_\nu-R\delta^\mu_\nu/2$ can be calculated from Eqs (\ref{7}) and (\ref{9}). We get
\begin{equation}
v_t=v_r=\frac{24 M(r-k)}{k^4},\>\>\>\>\>\>v_\theta=v_\phi=\frac{12 M(3 r-2 k)}{k^4}
\label{10}
\end{equation}
and the normalized eigenvectors  $v_{t\mu}$  and $v_{r\mu}$ corresponding to the eigenvalues $\lambda_t$ and $\lambda_r$ are
\begin{equation}
v_{t\mu}=\sqrt{1-\frac{8 M r^2}{k^3}+\frac{6 M r^3}{k^4}}\delta_{t\mu},\>\>\>\>\>v_{r\mu}=\frac{1}{\sqrt{1-\frac{8 M r^2}{k^3}+\frac{6 M r^3}{k^4}}}\delta_{r\mu}
\label{12}
\end{equation}

From Eqs (\ref{1}), (\ref{3}), (\ref{4}) and (\ref{9})-(\ref{12}) we get $ R_{\mu\nu}-\frac{R}{2}g_{\mu\nu}=8 \pi T_{\mu\nu}$, where
\begin{equation}
T_{\mu\nu}=(\rho+p_\perp)v_{t\mu}v_{t\nu}+p_\perp g_{\mu\nu}+(p_r-p_\perp)v_{r\mu}v_{r\nu}
\label{15}
\end{equation}
with
\begin{equation}
\rho=-\frac{v_t}{8 \pi}=\frac{3 M(k-r)}{\pi k^4}
\label{16}
\end{equation}
\begin{equation}
p_r= \frac{v_r}{8 \pi}=-\frac{3 M(k-r)}{\pi k^4}
\label{17}
\end{equation}
\begin{equation}
p_\perp=\frac{v_\theta}{8 \pi}=\frac{v_\phi}{8 \pi}=\frac{3 M(3 r-2 k)}{2 \pi k^4}
\label{18}
\end{equation}
Therefore the expression (\ref{1}) is an anisotropic fluid solution with energy density $\rho$, radial pressure $p_r$ and tangential pressure $p_\perp$ \cite{He1}

We easily find that
\begin{equation}
4\pi\int_0^k\rho r^2dr=M
\label{19}
\end{equation}
which means that the total mass in the interior region $0\leq r\leq k$ is $M$, as expected.

Eqs  (\ref{16})-(\ref{18}) give the relations $ \rho+p_r=0 $ and $ \rho+p_\perp=\frac{3 M r}{2 \pi k^4}$.
From these relations and Eq. (\ref{16}) we find that in the interior region $0\leq r\leq k$ the solution (\ref{1}) satisfies the weak energy conditions.

An equation of state of the form $\rho+p_r=0$ was introduced originally by Sakharov as an equation of state of a superdense  fluid \cite{Sa1}.
Gliner  \cite{Gl1} called the matter with this equation of state  $ \mu - vacuum $. He argues that the meaning of a negative pressure is that the internal volume forces in the matter are not forces of repulsion  but forces of attraction  and also that  an object with this equation of state might be formed in gravitational collapse.   This equation  arises in Grand Unified Theories at very high densities and it is used in the cosmological inflationary senario \cite{Li1}. Also it is the equation of state in the de Sitter interior of the gravastar model \cite{Ma3} \cite{Vi1}. It appears in many papers, in which regular black hole solutions are constructed \cite{An1}.

It is argued that the Einstein tensor $G_{\mu\nu}$ of a metric of the form  $ ds^2=-F(r)dt^2+\frac{1}{F(r)}dr^2 +r^2(d\theta^2+\sin^2\theta d \phi^2)$, where $F(r)$ has the asymptotic behavior $F(r)\sim1-\frac{r^2}{l^2}$ as $r\rightarrow0$, has the cosmological-constant form
\begin{equation}
 G_{\mu\nu}\sim-\Lambda g_{\mu\nu}\>\>\>\>\mbox{as}\>\>\>\>r\rightarrow0\>\> \>\>\>\mbox{with}\>\>\>\>\>\Lambda=\frac{3}{l^2}
\label{20}
\end {equation}
which means that we have an effective cosmological constant at small distances with Hubble length $l$ \cite{Ha1}. Since from Eqs (\ref{16})-(\ref{18}) we get $\rho+p_\perp\sim0$ and $p_r-p_\perp\sim0$ as $r\rightarrow0$ we find from Eq. (\ref{15}) that the above relation is satisfied for the metric of Eq. (\ref{1}) with $\Lambda=\frac{24M}{k^3}$. The same result for $\Lambda$ was found in Ref.  \cite{Dy1}, since $8\pi\rho\sim\frac{24M}{k^3}$

From Eq. (\ref{16}) we find that the energy density $\rho$ vanishes at the boundary $r=k$ and decreases monotonically as we move from the center to the boundary. Also $p_\perp$ and $\rho$ satisfy the relation $p_\perp=-\rho-\frac{r}{2}\frac{d\rho}{dr}$.

The curvature scalar $R^2$ of Eq. (\ref{1}) is given by the relation \cite{Bo1}

\begin{equation}
R^2=\frac{48 M^2}{k^8}(32 k^2-80 k r+57 r^2)
\label{22}
\end{equation}
Therefore the solution (\ref{1}) is regular since $R$ and $R^2$ are regular.

The horizons of the solution (\ref{1}) are obtained from the relation $g^{rr}=0$, from which we get
\begin{equation}
x^3-\frac{4}{3}x^2+w=0,\>\>\>\>where\>\>\>\>x=\frac{r}{k}\>\> \>\>\>\mbox{and}\>\>\>\>\>w=\frac{k}{6 M}
\label{23}
\end {equation}
If $x_i$ , where $ i=1,2$, are the real and positive roots of Eq. (\ref{23}) the horizons $r_{ih}$  are at $ r_{ih}=kx_i=6Mwx_i$.
Also Schwarzschild's solution has horizon at $ r_S=2M $, while the matching surface $ r_M $ is at $ r_M=k=6Mw$. Combining these expressions we get
\begin{equation}
r_M=3wr_S=\frac{r_{ih}}{x_i}
\label{28}
\end {equation}

Consider the cubic equation $x^3+a_2 x^2+a_1 x+a_0=0$ and define $q$ and $p$ by the relations $ q=\frac{1}{9}(3a_1-a^2_2)\>\>\>\mbox{and}\>\>\> p=\frac{1}{6}(a_1a_2-3a_0)-\frac{1}{27}a^3_2$.
Then if $q^3+p^2>0$ the solution has one real and two complex roots, if  $q^3+p^2=0$ all roots are real and at least two of them are equal and if  $q^3+p^2<0$ all roots are real \cite{Ab1}. In our case we find that $q^3+p^2>0$ if $w>w_0$ where  $ w_0=\frac{256}{729}=0,351166$,
$q^3+p^2=0$ if $w=w_0$ and $q^3+p^2<0$ if $w<w_0$.The solution $x=1$ of Eq. (\ref{23}) is obtained for $w=1/3$. We consider the cases:

(1) $w>w_0$: Since in this case two of the solutions are complex and the real solution is negative the regular solution we have found does not have an event horizon. Also from Eqs  (\ref{28}) we get
\begin{equation}
r_M > r_S
\label{31}
\end{equation}
Therefore the event horizon of Schwarzschild's solution is inside the matching surface. The combination of Schwarzschild's solution with its interior does not lead to the  formation of a black hole as the end product of gravitational collapse. It leads to the formation of a regular object with de Sitter center, a gravitational soliton, a G-lump, which is held by its gravitational self interaction. G-lumps are model independent gravastar ( gravitational vacuum stars ) and can be responsible for local  effects related to dark matter in a way similar to primordial black holes \cite{Dy3}.

The criterion of G-lump stability under extremal polar perturbations is given by \cite{Dy3}
\begin{equation}
\rho+(\rho+p_\perp)-r\frac{d(\rho+p_\perp)}{dr}\geq0
\label{32}
\end{equation}
The solution we have found satisfies the above expression in the interior region since $\rho+p_\perp=\frac{3Mr}{2\pi k^4}$ and $\rho\geq0$ in this region.

Since in all cases which follow one of the real solutions is negative this solution will  be ignored and for the other two solutions $x_1$ and $x_2$ we shall make the choice $ x_1 \geq x_2 $.

(2) $w=w_0$: In this case we have $x_1=x_2=8/9$ and
 we conclude that we have a " degenerate" event horizon at $ r_{1h} = r_{2h} =\frac{4096}{2187}M$. Also from Eqs (\ref{28}) we find that $ r_M>r_S>r_{1h} $.

(3) $w_0>w>\frac{1}{3}$: In this case we have $ r_M>r_S>r_{1h}>r_{2h}$

(4) $ w=1/3: $ In this case we find that $ r_M=r_S=r_{1h}>r_{2h}$.

(5) $w<\frac{1}{3}$: For these values of $w$ we get $ r_S>r_{1h}>r_M>r_{2h}$. \\
In cases (2)-(5) regular black hole appears in the final state of the collapse.

\end{document}